\documentstyle[aps,twocolumn,psfig]{revtex}
\begin{document}
\title{$U(2)$ invariant squeezing properties of  pair
coherent states}
\author{Arvind~\thanks{Electronic address:
arvind@physics.iisc.ernet.in}
}
\address{Department of Physics, \\
Guru Nanak Dev University, Amritsar 143005 India}
\maketitle
\begin{abstract}
The $U(2)\/$ invariant approach  is delineated  for the pair coherent
states to explore their squeezing properties. This approach is useful
for a complete analysis of the squeezing properties of these two-mode
states.  We use  the maximally compact subgroup $U(2)\/$ of
$Sp(4,R)\/$ to mix the modes, thus  allowing us to  search over all
possible quadratures for squeezing. The variance matrix for the pair
coherent states  turns out to be analytically diagonalisable, giving
us a handle over its least eigenvalue, through which we are able to
pin down the squeezing properties of these states.  In order to
explicitly demonstrate the role played by $U(2)\/$ transformations, we
connect our results to the previous analysis of squeezing for the pair
coherent states.
\end{abstract}
\section{Introduction}
Pair coherent states 
(originally discussed for the case of charged bosons~\cite{dutta-roy})
provide an interesting example of non-classical states of the two-mode radiation
field~\cite{agarwal-1,agarwal-2}.  They have been studied in detail for
their non-classical properties and as examples of  EPR
states~\cite{agarwal-1,agarwal-2,agarwal-3,reid-1,reid-2,munro-1}.
More recently, their experimental signatures have been
explored~\cite{munro-2}.

Squeezing is an important  signature of nonclassicality. By
definition, when 
the noise in some quadrature of a quantum state falls below the coherent state
value of $\hbar/2\/$, the state is squeezed and thus
non-classical~\cite{walls-nature-79,walls-nature-83}. The canonical
commutation relations which lead to the uncertainty principle are
fundamental to the analysis of squeezing. The linear canonical
transformations of quadrature operators, under which the canonical
commutation relations of the two-mode field are invariant, form a
non-compact group $Sp(4,R)\/$. The group $Sp(4,R)\/$ acts on the
quantum states of two-mode fields through the unitary
representation of its double cover and represents the action of
all possible quadratic Hamiltonians which are physically 
important~\cite{arvind-pra-1995}. The group $Sp(4,R)\/$ has a
passive, photon number conserving, maximally compact subgroup $U(2)\/$
which acts on the creation and annihilation operators through its
defining representation. The non-compact part of $Sp(4,R)\/$, while
acting through its unitary representation on  ``classical''
non-squeezed states can generate ``non-classical'' squeezed states.
On the other hand, the compact $U(2)\/$ subgroup of $Sp(4,R)\/$ through
its unitary action in the Hilbert space 
cannot generate non-classical states from classical
ones. In particular it cannot generate a squeezed state starting from a
non-squeezed state. Therefore, one can allow a state to undergo such a
passive transformation before analysing any non-classical
property~\cite{arvind-pra-1995,arvind-pla-1999}.  
This maximally compact subgroup $U(2)\/$ of $Sp(4,R)\/$ while acting
on the quadrature operators, allows us to search over all possible
allowed quadratures for the two-mode fields~\cite{arvind-pra-1995}.
This search enables one to locate the most squeezed
quadrature. Thus the group $U(2)\/$ facilitates the analysis of
quadrature squeezing and allows one to arrive at a $U(2)\/$ invariant
description of squeezing. It is possible to generalize this analysis
for the case of n-mode fields where the maximally compact subgroup of 
canonical transformations is $U(n)\/$~\cite{simon-pra-94}

In this paper, we analyse the squeezing
properties of pair coherent states 
using the $U(2)\/$ invariant methodology.  It turns out
that for these states the variance matrix (the matrix of all
second order noise moments) has an interesting form. We are able
to diagonalise it analytically through a series of orthogonal
transformations, thus locating its spectrum and hence the smallest
eigenvalue. The spectrum is invariant is under $O(4)\/$ and hence
also under $U(2)$ transformations.  Whenever this $U(2)\/$
invariant least eigenvalue is less than $\frac{1}{2}\/$ the state
is squeezed. Further, we show that the orthogonal $O(4)\/$ matrix
which diagonalises the variance matrix for these states lies outside
the compact canonical transformations (the subgroup $U(2)\/$). 
Therefore, the pair coherent
states provide an interesting example of states for which the
variance matrix cannot be diagonalised within $O(4) \cap Sp(4,R)\/$,
though we are able to bring the smallest eigenvalue to the leading
diagonal position by some compact canonical transformations
($O(4)\/ \cap \/Sp(4,R)\/$). Lastly we connect our results to the previous
analysis on the squeezing properties of pair coherent states by Agarwal et.
al.~\cite{agarwal-1,agarwal-2}

The material in this paper is arranged as follows: in
section~\ref{squeezing}  we
describe the $U(2)\/$ invariant squeezing criterion and elaborate on
the role played by $U(2)\/$ in the analysis of the non-classical
properties of the two-mode fields. In section~\ref{pair}  we analyse the pair
coherent states for their $U(2)\/$ invariant squeezing properties and
section~\ref{conclude}  contains a few  concluding remarks.
\section{$U(2)\/$ Invariant Squeezing }
\label{squeezing}
We consider two orthogonal modes of the radiation field, with
annihilation operators $a_1\/$ and $a_2\/$.  To handle the analysis of
the two-mode fields compactly  we introduce the column vectors
\begin{equation} 
\xi^{(c)} \equiv \left(\xi_a^{(c)}\right) =
\left(\begin{array}{c}
a_1\\a_2\\a^{\dagger}_1\\a^{\dagger}_2\end{array}\right),\quad\quad
\xi \equiv \left(\xi_a\right) = \left(\begin{array}{c}
q_1\\q_2\\p_1\\p_2\end{array}\right).
\end{equation} 

$\xi^{(c)}\/$  being the vector of creation and annihilation 
operators and $\xi\/$ the vector of the 
quadrature operators, with their
components having the usual relation,
$q_j=\frac{1}{\sqrt{2}}\left(a_j + a_j^{\dagger} \right)$ and
$p_j=\frac{-i}{\sqrt{2}}\left(a_j - a_j^{\dagger} \right)$.  The
canonical commutation relations can be  written compactly in 
terms of these column vectors:
\begin{equation} 
\begin{array}{ccc}
\mbox{}\left[\xi^{(c)}_a,\xi^{(c)}_b\right] &=&  \beta_{ab}\\
&&\\
\mbox{}\left[\,\xi_a^{},\xi_b^{}\,\right] &=& i
\beta_{ab} \end{array} \, \mbox{with}\quad\,
\left(\beta_{ab}\right) = 
\left( \begin{array}{rrcc}
\!\!0&\!\!\!\!0&1&0\\ \!\!0&\!\!0&0&1\\ \!\!-1&\!\!0&0&0\\
\!\!0&\!\!-1&0&0 \end{array} \right) 
\label{commutation}
\end{equation}

The linear canonical transformations of the quadrature operators
$q_j\/$ and $p_j\/$ are those real linear transformations that
preserve the commutation relations given in
equation~(\ref{commutation}). They
constitute the four-dimensional symplectic group $Sp(4,R)$: 
\begin{eqnarray} 
&\xi \longrightarrow \xi^{\prime} = S\, \xi\/, \quad
\quad S \in Sp(4,R)& \nonumber \\ 
&Sp(4,R) = \left\{ S= 4 \times 4
\mbox{ real matrix }\,\, \vert \,\, S\/ \beta \/S^{T}=\beta\right\}&
\end{eqnarray}
The maximally compact subgroup $K=U(2)\/$ of $Sp(4,R)\/$, which is
central to our  analysis of squeezing, can be identified as:
\begin{eqnarray}
&K= \left\{ S(X,Y) \in Sp(4,R) \,\, \vert \,\,  U=X-iY \in
U(2)\right\}&
\nonumber \\
&S(X,Y) = \left[
\begin{array}{rl} 
X &Y\\-Y&X \end{array}
\right]&
\end{eqnarray}
The action of this subgroup on the creation and annihilation operators
is through its defining representation:
\begin{equation}
\left[
\begin{array}{cc} a_1^{\prime}\\ a_2^{\prime} \end{array}
\right]
=U \left[\begin{array}{cc}a_1\\ a_2
\end{array}\right], \quad \quad  U \in U(2)
\end{equation}

The standard way of distinguishing classical from non-classical states
is through the diagonal coherent state
description~\cite{walls-nature-79,teich-progop-88}. A given
two-mode density operator $\rho\/$  can always be expanded
in terms of coherent states:
\begin{equation}
\rho = \int \frac{d^2z_1 d^2z_2}{\pi^2} \phi(z_1,z_2)
\vert z_1,z_2\rangle \langle z_1,z_2\vert
\label{phi}
\end{equation} 
where $\vert z_1,z_2 \rangle\/$ are the two-mode coherent states.
The unique normalized weight function $\phi(z_1,z_2)\/$
provides a complete description of the two-mode state $\rho\/$ and
can in general be a distribution which is quite
singular~\cite{klauder}.  For the case when
$\phi(z_1,z_2)\/$ can be interpreted as a probability
distribution (i.e. it is nonnegative and  nowhere more
singular than a delta function), equation~(\ref{phi})
implies that the state $\rho\/$ is a classical mixture of
coherent states which have a natural classical limit.
Such quantum states are referred to as ``classical''; in
contrast those states for which $\phi(z_1,z_2)\/$ either
becomes negative or more singular than a delta function,
are defined as  ``non-classical''.

When the two-mode state described by density operator $\rho\/$,
transforms under a unitary operator corresponding to the
compact $U(2)\/$ subgroup of $Sp(4,\Re)\/$, the
distribution $\phi(z_1,z_2)\/$ undergoes a point
transformation given in terms of the $U(2)\/$ element:
\begin{eqnarray}
&\rho^{\prime} = {\cal U}(S(X,Y)) \,\,\rho\,\, {\cal U}(S(X,Y))^{-1} 
\Longleftrightarrow& 
\nonumber \\
&\phi^{\prime}(z_1,z_2) = 
\phi(z_1^{\prime},z_2^{\prime}),&
\nonumber \\
&\left[\begin{array}{c} z_1^{\prime} \\ 
z_2^{\prime} \end{array} \right]
= U\left[\begin{array}{c} z_1 \\ z_2 \end{array}
\right]
,\,\,\,\,\, U=X-iY\in U(2)
&\!\!\!\!\!\!\!\!\!\!\!\!\!\!\!\!
\end{eqnarray} 
Thus, under $U(2)\/$ the classical states map onto
classical ones and the non-classical states onto non-classical ones; these
transformations are incapable of generating a non-classical state from
a classical one or vice versa.

We recapitulate and collect some interesting
and important properties of the maximally compact subgroup
$ K=U(2)\/$ of $Sp(4,R)\/$ here:
\begin{itemize}
\item[(a)] When $\xi^{(c)}$ undergoes a $U(2)\/$ transformation, the
annihilation operators $a_r$ are not mixed with the 
creation operators $a_r^{\dagger}$.  
\item[(b)] The action of the elements of $U(2)\/$ 
on a quantum state does not change the distribution of the total photon
number.
\item[(c)] The diagonal coherent state distribution
function is covariant under $U(2)\/$ transformations.
\item[(d)] One requires only passive optical elements to 
experimentally implement any $U(2)\/$ transformation on 
a state of the two-mode
field~\cite{yurke-pra-86,simon-pla-90,arvind-pra-1995}.
\end{itemize}

We see that the passive $U(2)\/$ transformations are a useful tool to
analyse the nonclassicality of a two-mode state.  Therefore, it is
reasonable to demand that any signature of non-classicality be
invariant under such transformations.  Using such transformations we
can always try to transform the nonclassicality if it is present but
hidden, into a more visible form.

We now turn towards the analysis of second order noise moments for a
general two-mode state and  $U(2)\/$ invariant description of
squeezing.  Let $\rho\/$ be the density operator of any (pure or
mixed) state of the two-mode radiation field. With no loss of
generality we may assume that the mean values
$\mbox{Tr}\,(\rho\xi_a)\/$ of $\xi_a\/$ vanish in this state (any
such non-zero values can always be reinstated by a suitable
phase space displacement which has no effect on the squeezing
properties). Squeezing involves the set of all second order noise
moments of the quadrature operators $q_j\/$ and $p_j\/$.  To manipulate 
them collectively we define the variance or noise matrix $V\/$ for the
state $\rho\/$ as follows:

\begin{eqnarray}
V & = & (V_{ab}),\nonumber\\ V_{ab} & = & V_{ba} =
\frac{1}{2}\mbox{Tr}\,(\rho\{\xi_a,\xi_b\}).
\label{V-def}
\end{eqnarray}
This definition is valid for a system with any number of
modes. For a two-mode system it can be written explicitly
in terms of $q_j\/$ and $p_j\/$ as:
\begin{equation}
V=\left(\begin{array}{cccc}
\langle q_1^2\rangle&\langle q_1 q_2\rangle &
\frac{1}{2}\langle \{q_1,p_1\}\rangle &
\langle q_1 p_2\rangle \\
\langle q_1 q_2\rangle &\langle q_2^2\rangle &
\langle q_2 p_1\rangle &
\frac{1}{2}\langle \{q_2,p_2\}\rangle \\
\frac{1}{2}\langle \{q_1,p_1\}\rangle &
\langle q_2 p_1 \rangle &
\langle p_1^2\rangle& \langle p_1 p_2 \rangle \\
\langle q_1 p_2\rangle&
\frac{1}{2}\langle \{q_2,p_2\}\rangle&
\langle p_1 p_2 \rangle & \langle p_2^2 \rangle
\end{array} \right)
\label{V-expr}
\end{equation}
This matrix is real symmetric positive definite and obeys
additional inequalities expressing the Heisenberg
uncertainty principle~\cite{simon-pra-94}. The four diagonal entries of
the variance matrix represent quadrature noise; of the six
independent off-diagonal entries, two are the expectation
values of the anticommutator between $q\/$ and $p\/$ of the
same mode while the remaining four represent mode-correlations.

When the state $\rho\/$ is transformed to a new state
$\rho^\prime\/$ by the unitary operator ${\cal U}(S)\/$ for
some $S\in Sp(4,R)\/$, we see easily 
that the variance matrix $V\/$
undergoes a symmetric symplectic or congruent transformation:
\begin{eqnarray}
S \in Sp(4,R): \rho^{\prime}  ={\cal U}(S)\;\rho\;{\cal
U}(S)^{-1}
\Rightarrow
 V^\prime  = S\;V\;S^T\,.
\label{vtransformation}
\end{eqnarray}
This transformation law for $V\/$ preserves all the
properties mentioned after equation~(\ref{V-expr}).

As has been discussed earlier and in detail
elsewhere~\cite{simon-pra-94},
for a multi-mode system it is
physically reasonable to set up a definition of squeezing
which is invariant under the subgroup of passive
transformations of the full symplectic group. For the
present case of two-mode systems, we evidently need a
$U(2)\/$-invariant squeezing criterion. Our
definition must be such that, if a state $\rho\/$ with
variance matrix $V\/$ is found to be squeezed(non-squeezed),
then the
state $ \rho^{\prime}={\cal U}(S(X,Y))\;\rho\;{\cal U}(S(X,Y))^{-1}\/$
with variance matrix $V^\prime = S(X,Y)\;V\;S(X,Y)^T\/$ must
also be squeezed(non-squeezed), for any $U = X-iY\in U(2)\/$.
(where ${\cal U}(S(X,Y)\/$ is the unitary operator corresponding to 
$S(X,Y)\/$).
Conventionally, a state is said to be squeezed if any one of
the diagonal elements of $V\/$ is less than 1/2 (we are working with
$\hbar=1$). The
diagonal elements correspond, of course, to fluctuations in
the ``chosen'' set of quadrature components of the system.
The $U(2)\/$-invariant definition is as follows: the state
$\rho\/$ is a quadrature squeezed state if either some
diagonal element of $V\/$ is less than 1/2 (and then we say
that the state is manifestly squeezed), or some diagonal
element of $V^\prime = S(X,Y)\;V\;S(X,Y)^T$ for some
$U=X-iY\in U(2)\/$ is less than 1/2:
\begin{eqnarray}
\lefteqn{\rho\;\mbox{is a squeezed state}\;\Leftrightarrow}\nonumber\\
&&\left(S(X,Y)\;V\;S(X,Y)^T\right)_{aa} < \frac{1}{2}, 
\mbox{(no sum over $a\/$} \nonumber)\\
&&\mbox{for some $a\/$ and some}\; X-iY\in U(2). 
\label{sq-crt}
\end{eqnarray}
Thus searching over $S(X,Y)\in U(2)\/$ is synonymous
with  exploring all possible sets of quadrature components.
We may say that since any element of $U(2)\/$ passively
mixes the two modes, the appropriate $S(X,Y)\in U(2)\/$
which achieves the above inequality (assuming the given
$V\/$ permits the same) just chooses the right combination
of quadratures to make the otherwise hidden
squeezing manifest.

To implement this definition in practice, it would appear
that even if a state is intrinsically squeezed, we may have
to explicitly find a suitable $U(2)\/$ transformation which
when applied to $V\/$ makes the squeezing manifest. This
however could be complicated. The point to be noticed
and appreciated here is, that diagonalisation of a noise matrix
$V\/$ generally requires a real orthogonal transformation
belonging to $SO(4)\/$, which may not lie in $U(2) = O(4)
\cap Sp(4,R)$. It is therefore remarkable
that, as shown in \cite{simon-pra-94}, the $U(2)\/$-invariant
squeezing criterion (\ref{sq-crt}) can be expressed in
terms of the spectrum of eigenvalues of $V\/$, namely:
\begin{equation}
\mbox{$\rho\/$ is squeezed}\;\Leftrightarrow\; 
\mbox{least eigenvalue of}\;V < \frac{1}{2}.
\label{sq-crt-mat}
\end{equation}
While the diagonalisation of $V\/$ is in general
not possible within $K = U(2)\/$ which is a proper subgroup
of $O(4)\/$ any one particular (and hence the smallest)
eigenvalue of $V\/$ can be brought to the leading diagonal position
of $V^{\prime}\/$ obtained from $V\/$ by
transformation through an appropriate
$S(X,Y) \in U(2)$. In other words any quadrature component
can be taken to any other quadrature component by a
suitable element of $U(2)$. We shall henceforth work
with the $U(2)\/$-invariant squeezing criterion given in equations
(\ref{sq-crt}) and~(\ref{sq-crt-mat}) and analyse the pair coherent states 
from this point of view in the next section.
\section{Pair Coherent States}
\label{pair}
We now turn to the pair coherent states $\vert \zeta,
q \rangle \/$ defined as
\begin{eqnarray}
a_1 a_2 \vert \zeta, q \rangle = \xi \vert \zeta, q \rangle,\quad
(a_1 a_1^{\dagger} -a_2 a_2^{\dagger}) 
\vert \zeta, q \rangle = q \vert \zeta, q \rangle
\end{eqnarray}
These are non-classical states with fixed number of photon difference
$q\/$ between the two modes.  The other eigenvalue $\zeta\/$ is in
general complex.
The solution to this eigenvalue equation was found by Bhaumik et.
al.~\cite{dutta-roy} in the context of charged bosons.
This eigenvalue problem was studied in the context of 
competitive nonlinear effects in two-photon media by
Agarwal~\cite{agarwal-1,agarwal-2} and the
important  quantum character of fields in such states was  also
pointed out by him.

Assuming $q\/$ to be positive the solution to the eigenvalue problem
turns out to be
\begin{equation}
\vert \zeta, q \rangle =N_q \sum^{\infty}_{n=0} \frac{\zeta ^n}{[n!
(n+q)!]^{1/2}}\/\/\vert
n+q,n\rangle
\end{equation}
where the normalisation constant $N_q\/$ is given by
\begin{equation}
N_q = \left[\sum^{\infty}_{n=0} \frac{\vert
\zeta\vert^{2n}}{n! (n+q)!} \right]^{-1/2}
\end{equation}
For our analysis we first calculate all the
noise moments for this family of states, which becomes easy if we
first compute
the photon numbers in the two modes involved. The photon number in the two
modes $N_1\/$ and $N_2\/$ is given by
\begin{eqnarray}
N_2 &=&\langle a_2^{\dagger}a_2\rangle = 
\left[\frac{N_q}{N_{q+1}}\vert \zeta \vert \right]^2
\nonumber \\
N_1 &=&\langle a_1^{\dagger}a_1\rangle = 
N_2+q
\label{numbers}
\end{eqnarray}
Now turning to the quadrature components, as usual we find that the
first  moments vanish
\begin{equation}
\langle q_1\rangle=
\langle q_2\rangle=
\langle p_1\rangle=
\langle p_2\rangle=0
\end{equation}
The second moments can also be readily evaluated using the
equation~(\ref{numbers})to give us the
variance matrix defined in equation~(\ref{V-expr}) for the state
$\vert \xi, q  \rangle\/$ as
\begin{equation}
V(\zeta,q)=\left[\begin{array}{cccc}
N_1+\frac{1}{2} &\mbox{Re} \/\zeta & 0&\mbox{Im}\/ \zeta \\
\mbox{Re}\/ \zeta& N_2+\frac{1}{2} & \mbox{Im}\/ \zeta &0\\
0 & \mbox{Im}\/ \zeta &N_1+\frac{1}{2} & - \mbox{Re}\/ \zeta\\
\mbox{Im}\/ \zeta  &0 &-\mbox{Re}\/ \zeta & N_2+\frac{1}{2}
\end{array}
\right]
\end{equation}

For the analysis of squeezing properties we need to calculate the
smallest eigenvalue of the  above matrix which we do by explicit
analytical diagonalisation. It is convenient to
rewrite $V(\zeta,q)\/$ as a sum of two matrices, a multiple of  
identity and a  matrix with a block form as follows:
\begin{equation}
V(\zeta,q)=(N_2+\frac{1}{2})I +
\left[\!\!
\begin{array}{cc} 
\left[ \begin{array}{cc} q &\mbox{Re}\/ \zeta \\ 
 \mbox{Re}\/ \zeta & 0 \end{array}\right]
 & \mbox{Im}\/ \zeta 
\left[\begin{array}{cc} 0&1 \\ 1&0
\end{array}\right]\\
&\\
{\mbox{Im}\/ \zeta \left[\begin{array}{cc} 0&1\\1&0\end{array}\right]}&
{\left[\begin{array}{cc} q &-\mbox{Re}\/ \zeta \\ 
                  -\mbox{Re}\/ \zeta& 0 \end{array}\right]}
\end{array}\!\!
\right]
\label{pair-var}
\end{equation}
The structure of the block form suggests that  the diagonalisation
can be achieved in a two step process.
First we implement a rotation which diagonalises the diagonal blocks
and keeps the off diagonal blocks invariant. This can be achieved by
a rotation matrix of the following block diagonal form:
$$
R_1(\theta)=\left[
\begin{array}{cc}
R(\theta) & 0 \\
0 &R(\theta)^{T}
\end{array}\right]
$$
Here, $R(\theta)\/$ is the $2\times2 \/$ rotation matrix such that
the matrix
$$
R(\theta) 
\left[ \begin{array}{cc} q &\mbox{Re} \zeta \\ 
 \mbox{Re} \zeta & 0 \end{array}\right] R(\theta)^{T} 
$$
is diagonal. The lower diagonal part of $R_1(\theta)\/$ matrix is chosen to be
$R(\theta)^T\/$ so as to diagonalise the corresponding lower $2\times 2\/$
block in equation~(\ref{pair-var})
After such a rotation the variance matrix becomes
\begin{eqnarray}
V(\zeta,q)^{\prime}&=& R_1(\theta) V(\zeta,q) R_1(\theta)^{T}
\nonumber \\
&=&(N_2+\frac{1}{2}) I +
\left[
\begin{array}{cccc}
\lambda_{+} & 0& 0 & \mbox{Im} \zeta \\
0&\lambda_{-} & \mbox{Im} \zeta &0\\
0& \mbox{Im} \zeta &\lambda_{+}&0\\
\mbox{Im} \zeta & 0& 0&\lambda_{-}
\end{array}\right]
\nonumber \\
\mbox{with}&&\quad
\lambda_{\pm}=\frac{1}{2}\left(q\pm\sqrt{q^2+4\/\mbox{Re}}\/
\zeta\right) 
\end{eqnarray}
The diagonal blocks have been diagonalized while  
the off diagonal blocks remain the same.

The matrix $V(\zeta,q)^{\prime}\/$ is block diagonal in terms of
the $1-4\/$ and $2-3\/$ subspaces. Thus the problem has been reduced to
two, $2\times 2\/$ diagonalisation problems which are easily tractable
by a  rotation matrix implementing independent rotations in  the
$1-4\/$ and $2-3\/$ subspaces, given as
\begin{equation}
R_2(\phi)=\left[\begin{array}{cccc}
\cos \phi & 0&0& \sin \phi\\
0 & \cos\phi & \sin \phi & 0\\
0 & -\sin\phi & \cos \phi & 0\\
-\sin\phi & 0 & 0& \cos\phi
\end{array}\right]
\end{equation}
with the value of the parameter $\phi\/$ chosen such that the matrix
\begin{equation}
\left[
\begin{array}{cc} 
\cos \phi & \sin\phi\\
-\sin\phi & \cos\phi
\end{array}\right]
\left[\begin{array}{cc} 
\lambda_{+} &\mbox{Im} \zeta\\
\mbox{Im}\zeta &\lambda_{-} 
\end{array}
\right]
\left[
 \begin{array}{cc}
 \cos \phi & -\sin\phi\\
 \sin\phi & \cos\phi
\end{array}\right]  
\end{equation}
is diagonal.

This process readily gives us the eigenvalues of the 
matrix $V(\zeta,q)\/$ which are doubly degenerate 
\begin{eqnarray}
e_{\downarrow}=N_2+\frac{1}{2}+\frac{q}{2} - \frac{1}{2}\sqrt{\left(q^2+4
\vert \zeta \vert^2 \right)}   
\nonumber \\
e_{\uparrow}=N_2+\frac{1}{2}+\frac{q}{2} + \frac{1}{2}\sqrt{\left(q^2+4
\vert \zeta \vert^2 \right)}   
\label{eigen}
\end{eqnarray}
where $e_{\downarrow}\/$ is the smaller eigenvalue. For a given
value of $\zeta\/$ and $q\/$, if the smaller eigenvalue turns out to
be less than $\frac{1}{2}\/$ the state is squeezed.

\begin{figure}
\label{eigen-value}
\hspace*{-5cm}
\psfig{figure=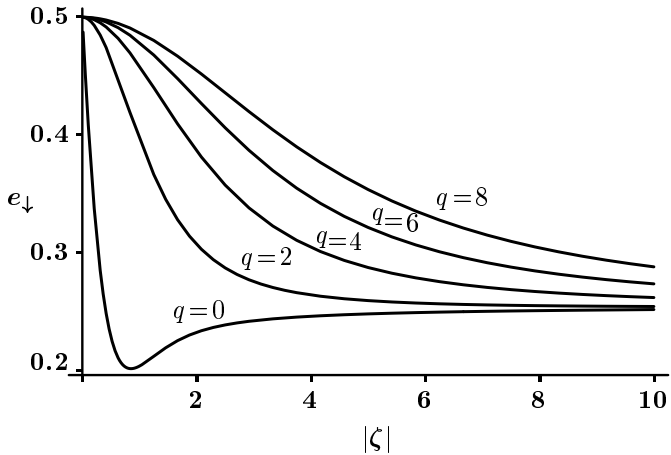}
\vspace*{-18cm}
\caption{Plot of the least eigenvalue of the variance matrix for the
pair coherent states as a function of $\vert zeta \vert\/$ for
different values of $q\/$, demonstrating the $U(2)\/$ squeezing for
these states.}  
\end{figure}

For example for $q=0\/$, and for small $\vert \zeta\vert\/$ (so that we can
neglect $N_2 \sim \vert
\zeta\vert^2\/$ compared to $\vert \zeta\vert$) we have 
\begin{equation}
e_{\downarrow} = \frac{1}{2} - \vert \zeta \vert \,<\,
\frac{1}{2}
\end{equation}
and therefore the state is squeezed independent of the phase of
$\zeta\/$.
For other values of $q\/$ and $\zeta\/$ we can determine the squeezed
or non-squeezed nature of the state by substituting the values of $q\/$
and $\zeta\/$ in expression for the least eigenvalue
$e_{\downarrow}\/$ in the equation~(\ref{eigen}). The plots of the
least eigenvalue as a function of $\vert \zeta\vert\/$ for different
values of $q\/$ are shown in Figure~(\ref{eigen-value}). The plots
reveal the squeezed nature of the states for a wide range of
parmeters. 
\newpage
\vspace*{-4cm}
\mbox{}

A closer look at the rotation matrices  $R_1(\theta)\/$ and
$R_2(\theta)\/$ reveals that while $R_2(\theta)\/ \in O(4) \cap
Sp(4,R)\/$, the matrix  $R_1(\theta)\/$ is an $O(4)\/$ transformation
not in $Sp(4,R)\/$ because it cannot be written in the $S(X,Y)\/$
form.  Thus the diaganalisation of the variance matrix $V(\zeta,q)\/$
is not possible within  $U(2)\/$.  The analysis in the preceding
section shows that even though we are not able to diagonalise the
variance matrix using compact canonical transformations we are
fortunately  able to bring the least eigenvalue to the leading
diagonal position using such transformations.  Physically this means
that a new set of quadratures can be arrived at through some
$S(X,Y)\/, \,\, U=X-iY \in U(2)\/$ such that one of them has its
second noise moment equal to the least eigenvalue $e_{\downarrow}\/$
of the variance matrix $V(\zeta,q)$. To demonstrate this explicitly,
consider a specific $U_{\psi} \in U(2)\/$ and the corresponding
transformation $S(X,Y)\/$:

\begin{eqnarray}
&U_{\psi} = X_{\psi} -iY_{\psi} =\frac{1}{\sqrt{2}} 
\left[\begin{array}{cc}
e^{-i\psi/2}&e^{-i\psi/2}\\ e^{i\psi/2} & -e^{i \psi/2}  
\end{array}  \right]&\nonumber\\
&X_{\psi} =\frac{1}{\sqrt{2}}\left[ \begin{array}{cc}
\cos{\frac{\psi}{2}} &
\cos{\frac{\psi}{2}}\\cos{\frac{\psi}{2}} & -\cos{\frac{\psi}{2}}
\end{array}\right]&\nonumber \\
&Y_{\psi} =\frac{1}{\sqrt{2}}\left[ \begin{array}{cc} 
+\sin{\frac{\psi}{2}} & +\sin{\frac{\psi}{2}} \\ 
-\sin{\frac{\psi}{2}} & \sin{\frac{\psi}{2}}
\end{array}\right]&\nonumber \\
&S(X_{\psi} ,Y_{\psi} )=\frac{1}{\sqrt{2}}\left[ \begin{array}{rr} X_{\psi}  &
Y_{\psi} \\ -Y_{\psi} &X_{\psi}  \end{array}\right].&
\label{h-class}
\end{eqnarray}
Using the matrix $S(X_\psi,Y_\psi)\/$ we can obtained the
transformed variance matrix
\begin{equation}
V^\prime(\zeta,q)=  
S(X_\psi,Y_\psi) V(\zeta,q)  S(X_\psi,Y_\psi)^{T}
\end{equation}
For a particular choice of 
$U_\psi\/$ with $\psi\/$ chosen to be the phase of $\zeta\/$, the transformed
matrix $V^{\prime}(\zeta,q)\/$ has its least eigenvalue in the
leading diagonal position. Therefore, for this particular value of
$\psi\/$ the compact transformation $U_{\psi}\/$ gives us the
new set of quadratures with one of them being the most squeezed one.
\section{concluding Remarks}
\label{conclude}
In this paper we have given a $U(2)\/$ invariant description of
squeezing properties of the pair coherent states.  For the pair
coherent states the variance matrix is not diagonalisable through
transformations within $U(2)\/$ and we have to use the full $O(4)\/$ 
for its diagonalisation. Therefore, these states  provide an
interesting example where the results of equation~(\ref{sq-crt-mat}) 
has to be
really used. We locate the least eigenvalue of the variance matrix 
$V(\zeta,q)\/$ by its explicit analytical diagonalisation
through an $O(4)\/$ transformation which is outside
$U(2)=O(4)\cap Sp(4,R)\/$, but the results of equations~(\ref{sq-crt})
and~(\ref{sq-crt-mat}) ensure that we can  also  bring this smallest
eigenvalue to the leading position (without diagonalising the
variance matrix) by a suitable $U(2)\/$ transformation. We have constructed
such a $U(2)\/$ transformation explicitly. This demonstrates the
power and elegance of the group theoretic approach.

The  smallest eigenvalue of the variance matrix
$V(\zeta,q)\/$  turns out to be independent of the phase of $\zeta\/$,
while the explicit $U(2)\/$ transformation which brings the least 
eigenvalue to the leading position depends upon the phase of
$\zeta\/$. This shows that as we scan through states with different
values for the phase of $\zeta\/$ we have to use  different $U(2)\/$
transformations to locate the maximally squeezed quadrature. In fact the
quadrature which is maximally squeezed is different for the
states related  to  each other by  such a phase factor.

In the usual analysis of squeezing one restricts oneself  to a
particular quadrature or a set of quadratures related to each other by
certain subclass within $U(2)\/$. Normally this subclass is dictated
by the measurement scheme being used. For the two-mode case for
example, the usual analysis considers the subclass which can be explored
through the heterodyne detection scheme. It is interesting to note 
that for the pair coherent states it
indeed suffices to consider the heterodyne subclass and the 
set of transformations given in equation~(\ref{h-class}) are exactly
those which can be explored using the heterodyne detection scheme~\cite{agarwal-1}. 
This is in general not true and we must consider the group $U(2)\/$ in
its entirety as a means to explore all the allowed quadratures for the two-mode
fields.

\noindent
{\bf Acknowledgement:}
The author thanks N. Mukunda and R. Simon for useful discussions.

\end{document}